# Sociological and Communication-Theoretical Perspectives on the Commercialization of the Sciences

Loet Leydesdorff

Amsterdam School of Communication Research (ASCoR), University of Amsterdam, Kloveniersburgwal 48, 1012 CX Amsterdam, The Netherlands;

loet@leydesdorff.net; http://www.leydesdorff.net

**Abstract**

Both self-organization and organization are important for the further development of the sciences: the two dynamics condition and enable each other. Commercial and public considerations can interact and "interpenetrate" in historical organization; different codes of communication are then "recombined". However, self-organization in the symbolically generalized codes of communication can be expected to operate at the global level. The Triple Helix model allows for both a neo-institutional appreciation in terms of historical networks of university-industry-government relations and a neo-evolutionary interpretation in terms of three functions: (*i*) novelty production, (*i*) wealth generation, and (*iii*) political control. Using this model, one can appreciate both subdynamics. The mutual information in three dimensions enables us to measure the trade-off between organization and self-organization as a possible synergy. The question of optimization between commercial and public interests in the different sciences can thus be made empirical.

**Introduction**

As an introduction to this topical issue of the "philosophy of commercialized science," the guest editor has listed a number of broad questions at the philosophical level. My own interest in STS (science & technology studies) and more specifically in the Triple Helix of university-industry-government relations is empirically oriented. However, I agree that the Triple Helix thesis (Etzkowitz & Leydesdorff, 2000) relates to philosophical and ethical issues such as the differences between commercialization and entrepreneurship in university-industry relations (Etzkowitz, 2002; Mirowski & Sent, 2007; Slaughter & Rodes, 2004). Furthermore, the commercialization of quantitative science studies and the development of bibliometric indicators has driven this debate home because of the use of indicators for legitimatory purposes (e.g., Bornmann *et al.*, 2008; Husbands Fealing *et al.*, 2011; Leydesdorff, 2008; Mirowski, 2010). The interest of journals and publishing houses in improving their "impact factors" or institutions in order to improve their rankings function increasingly as a research incentive in quantitative STS.

In what follows, I develop a sociological perspective inspired by the writing of Talcott Parsons (e.g., 1968) and Niklas Luhmann (e.g., 1990a-c). The focus is on their notion of symbolically generalized media of communication: how are scientific discourses codified differently from economic exchange processes or political power relations, and how strongly can such codifications shield domains against differences among domains? I shall argue that codification in interhuman communication does not "close" the discourse or exchange relations operationally, as Luhmann (1995a) argued using a metabiological metaphor (cf. Kuhn, 1962; Maturana, 1978). Cultural codifications, on the contrary (but under specific conditions), enable us precisely to generate "immutable mobiles" (Latour, 1987) within translation processes (Callon & Latour, 1981). The concept of hybrid networks can thus be elaborated in terms of networks in which different codes of communication can interact, and potentially generate synergies.

When symbolic generalization ("translation" into "immutable mobiles") is successfully achieved,[1] diffusion processes in domains other than that of the laboratory can be expected because one no longer needs to "unpack" the black-boxed instruments before using them (Callon, 1998). In an empirical study about the emergence and diffusion of two new techno-scientific developments—"small interference RNA" (siRNA) and "nanocrystalline solar cells" (NCSC)—Leydesdorff & Rafols (2011), for example, found the difference in terms of socio-cognitive "translatability" across disciplinary borders among specialties to be a crucial factor in moving the development from a disciplinarily oriented Mode-1 type of research to a transdisciplinary Mode-2 regime (Gibbons *et al.*, 1994).

One can consider such a transition from a local trajectory during the construction phase towards a globalizable regime during the diffusion phase as an "obligatory passage point" for agents (Callon, 1986) or a historical path-dependency from an evolutionary perspective (David, 1985). In the initial phase, new entrants attach preferentially to the inventor(s) who function as senders of the innovative information. In the later phase, newcomers can be expected to attach preferentially to the "centers of excellence" in which they wish to be received. Such centers can develop oligopolistically a division of labour in the techno-markets of future market potentials and government subsidies.

---

[1] This symbolic generalization of a specific instance is not yet symbolic generalization at the level of codes of communication.

The change in dynamics in these techno-science-market developments perhaps resembles more closely Schumpeter's (1943) shift from Mark-I to Mark-II (Freeman & Soete, 1997) than the notion of a global (that is, system-wide) transition from Mode-1 to Mode-2 research that has hitherto prevailed in science-policy discussions (Hessels & Van Lente, 2008). The process of bottom-up construction can gradually become inverted as the overlay of communications and exchanges above the level of local activities gains momentum and develops its own dynamics. This emerging overlay functions as a receptive feedback (that is, selection mechanism) on the communication dynamics among the senders. Whereas this additional layer of communication may initially add to the uncertainty which prevails, it can be expected to drift (in terms of selection mechanims operating and resonating) towards another steady-state as its attractor and then begin to reduce the prevailing uncertainty as a new regime is increasingly shaped.

The feedback from the emerging next-order regime tends to take over control from bottom-up trajectories. However, the "top-down" dynamics remains a socio-cognitive construct and therefore fragile, uncertain, and under reconstruction. In my opinion, Luhmann's (e.g., 1995a) socio-cybernetic ideas about the structural coupling of interhuman communication to human agency provide a useful heuristic for the understanding of these evolutionary mechanisms. Furthermore, operationalization can be envisaged using methodologies based on Shannon's (1948) mathematical theory of communication (Krippendorff, 1994, 2009; Leydesdorff, 2010a, 2011).

**Theoretical background**

The evolutionary metaphor of paradigms and paradigm changes (Kuhn, 1962) has revolutionized the empirical study of science and technology as socio-cognitive constructs (Barnes & Dolby, 1970; Barnes & Edge, 1982). Bloor (1976) notably translated these notions into the "Durkheimian paradigm" of the sociology of scientific knowledge (SSK): scientific knowledge can be considered as "social imagery" of the carrying communities of agents (Douglas, 1982). Slezak (1989) noted that from this perspective, the larger (cognitive) variance is explained in terms of the smaller (social) one. Discourse analysis (Mulkay *et al.*, 1983) and the French sociology of translation, however, have deconstructed agency in terms of communication processes, but instead of reversing the arrow for the explanation, the very objective of explanation has been abandoned by these authors in favour of an ethnographic approach (e.g., Latour, 1988 and 1992; Leydesdorff, 2007; Woolgar & Ashmore, 1988).

In my opinion, scientific discourse can be considered as highly-codified communication among intentional human beings; the specific codification (the jargon) enables us to entertain a dynamics in addition to belief systems, namely, as the communication of rationalized expectations. Unlike beliefs that can be attributed to agents or communities of agents, rationalized expectations are attributes of the discourse. They can be entertained and then partially reconstructed by agency. With reference to Husserl's ([1935/6], 1962) phenomenology, Luhmann (1990b, 2002a) emphasized that these *constructed constructs* rather than the *active constructing* were to be considered as the driving forces of social evolution.

Luhmann (1995) took from Maturana (e.g., 1978) the notion of "operational closure" of the communication and associated this with Kuhn's (1962) notion of "closed" paradigms (cf. Habermas, 1987; Leydesdorff, 2000). More than other sociologists, Giddens (1976) pointed to the problem of thus favouring concordance in belief structures over discourse when he wrote (at p. 143f.):

> "exaggeration of the internal unity of paradigms means that the latter tend to be treated by Kuhn as 'closed' systems. This leads to a characterictic difficulty in dealing with meaning-variance between paradigms […]. How is it possible to get from one meaning-frame to another, if they are separate and self-enclosed universes? […] In lieu of this, we must substitute, *as a starting-point*, that *all paradigms* (read 'language-games', etc.) *are mediated by the others*. This is so both on the level of the successive development of paradigms within science, and of the actor's learning to 'find his way about' within a paradigm."

A similar notion can also be found as an undercurrent in the work of Niklas Luhmann (1977, 1995b, 2002a) when he argues that reflexivity in the communication of meaning allows for "interpenetration" among meaning-processing systems (human agents and interhuman communications). In his philosophy of society, "interpenetration" is considered as an "operational coupling" in the reflexive communication of meaning (Luhmann, 1995b, p. 51; 2002a, p. 182) which adds to the "structural coupling" between human consciousness and communication. The structural coupling between consciousness at the individual level and communication at the above-individual level first closes these two domains operationally. Such closure, for example, enables agents to feel responsibility as "persons" who can act autonomously. Interpenetration as operational coupling among the meaning processing within the two thus distinguished systems enables us to translate locally—that is individually or in organizations—among differently codified communications (Luhmann, 1995b, p. 153).

I know that *operational closure* is more commonly emphasized when reading Luhmann (e.g., Luhmann, 2002b; cf. Leydesdorff, 2010b)—instead of this *operational coupling* by way of reflexivity. Operational closure is (meta-)biologically implied in Maturana's (1978) original concept of structural coupling. Künzler (1987) already noted that Luhmann deviated from Parsons by considering coding from the perspective of the biological metaphor of DNA and not sufficiently in terms of language as mediation. In my opinion, one should read Parsons' original idea of symbolic generalization of the codes of communication together with Luhmann's elaboration that these functionally different codes can be expected to span an architecture in the network along eigenvectors of this network. The differentiation in terms of these latent dimensions can be expected to enhance the communication of meaning by symbolically coding from different perspectives. More complexity thus can be processed, and at higher speeds.

However, differentiation and integration are two sides of the same coin. Under empirical conditions the one dynamics may prevail over the other. A high culture, for example, can be considered as mainly an integrated society, such as ancient Egypt or perhaps even 18th-century France. The symbolic integration of France in the anointed body of the King made it necessary to remove his head before a differentiated system guided by Montesquieu's *trias politica* could be constructed as a Constitution. Political discourse thus became institutionalized as a coordination mechanism of society alongside and in interaction with exchange processes on the market. This social construction in terms of *two* main coordination mechanisms led historically to a political economy.

A political economy is possible when exchange relations on the market and political communication provide two degrees of freedom for the social coordination. A third coordination mechanism prevailing at the level of society since the 19th century—that is, since the techno-scientific revolution of the period 1870-1910 (Braverman, 1974, Noble, 1977)—has been the socially organized production of knowledge (Whitley, 1984) in research laboratories both in industrial and state-financed settings. The additional option allows for the gradual transition from a political economy to a knowledge-based economy (David & Foray, 2002; Foray, 2004; Leydesdorff, 2006).

Whereas the codes of the communication are different—for example, one cannot buy "truth" with money—and therefore tend toward differentiation, integration implies reduction of complexity by interfacing and translating among the media in historical contexts. Thus, a specific configuration is shaped and reproduced. From the evolutionary or cybernetic perspective, this historical configuration of agency and institutions can be considered as the retention mechanism of a non-linear communication dynamics which further develops in terms of the flows of communications through the networks over time.

Luhmann expressed this trade-off as a kind of dialectics between *self-organization* of the "free" market forces, and other civil liberties, versus the "enabling and constraining" conditions of historical *organization* (Giddens, 1984). To this dialectics—between self-organization and organization—he added a third mechanism: "interactions". However, in my opinion, he distinguished insufficiently between interactions among human agents (e.g., in face-to-face communications) and interactions among communications as different units of analysis. The latter generate the variation in the social system of meaning-processing.

Elsewhere, I elaborated these three mechanisms—interaction, organization, self-organization—using three different mathematical formalizations of the anticipations implied (Leydesdorff, 2008, 2010c). In this context, I need to introduce from that discussion only the difference between historical and evolutionary mechanisms. Historical mechanisms are "phenotypically" observable, whereas evolutionary mechanism can be considered "genotypical". In philosophical terms, one could thus wish to extend the phenomenology of observables (Husserl, 1929, 1935/6; Schütz, 1932) with what one might call a more analytical "genomenology" of latent dimensions in interhuman communications. The "genotypes" in this case can be considered as the symbolically generalized codes in the expectations operating when meaning is communicated. The reference to Husserl here is not incidental because meaning is always communicated with reference to horizons of meaning, that is, other possible meanings. Meaning-processing thus can be expected to open up for other possible options when the various meanings can interact.

This proliferation and generation of redundancy—more possibilities—can further be reinforced by using models as in discursive knowledge (Giere, 1992 and 2006). Models enable us to select and codify specific meanings among possible meanings, and thus to specify expectations. The expectations are about future states which can reflexively be made operational in the present. This use of future states in the present tends to invert the historical arrow of time locally (Coveney & Highfield, 1990; Dubois, 1998; Prigogine & Stengers, 1988). However, this reflexive inversion operates in history and therefore leaves an empirically contingent footprint.

One should not reify "self-organization" as Greek gods who intervene in our lives. The models remain constructs, but the codified and symbolically generalized character of the communications can change their status from historical mechanisms—that is, at the level of trajectories—to evolutionary mechanisms operating at the level of regimes (Dosi, 1982). In *Capital I*, for example, Marx (1867) went to great lengths to explain why the exchange of commodities by the mediation of money ("Ware-Geld-Ware" or WGW) is different from the exchange of money via the mediation of commodities ("Geld-Ware-Geld" or GWG).² The latter mechanism is more abstract and thus able to take over evolutionary control under the condition of modern capitalism.

Similarly, in science—albeit perhaps less frequently than the daily transactions on the market—the submission of a knowledge claim (usually in the form of a manuscript) brings the context of justification into operation within the context of discovery in which this new insight had previously been entertained. The epistemological status of the knowledge content under discussion is thus changed. From a communication-theoretical perspective (cf. Popper, [1935] 1959), the "context of justification" remains socially contingent (composed, for example, of editors, referees, publishing houses, etc.), but the medium of communication has changed, and accordingly the rules of the game are changed. Note that the one subdynamics does not replace the other, but control may shift from the one to the other level, and back! The context of discovery remains needed for providing the variation, but it operates with a speed different from the context of justification (Simon, 1973).

In my opinion, this conceptualization in terms of different dynamics operating upon each other opens a realm of empirical questions. The symbolically generalized codes of communication function as auto-regenerative mechanisms because they are anchored in a multitude of dimensions in interhuman communication. Codification operates by deselecting most modalities and enhancing only one or a few. For example, education is codified differently from research. Political discourse is different from scholarly discourse and one is not supposed to pay one's way into a scholarly publication. A submission fee, for example, can be expected to have more symbolic meaning than value.

---

² "Ware-Geld-Ware" can be translated as "Commodigy-Money-Commodity" (Marx, 1995, pp. 66 ff.).

However, the digital turn in communication, which allows for entrepreneurship beyond the printing of journals, has witnessed more recently the emergence of full author fees as financial coverage for publication costs (Wyatt, 2010). If all journals were author-fee-based—against subscription-based—the science system as a self-organizing system might become driven exclusively by institutional and commercial concerns. Since not everyone is able to pay such a fee, control over access could shift to mechanisms other than self-regulation at the level of the scientific community (Wagner, 2008).

The question of this theme issue is, from this perspective, whether such a change at the control level has indeed occurred or might be occurring at present. From my structuralist perspective, I would not expect this shift to prevail in the longer run because differentiation and self-organization allow for greater complexity of the communication, whereas control and integration require organization from a perspective that reduces complexity (e.g., in terms of management objectives). But under historical conditions one may find situations where integrating organization (e.g., state or industrial control) is dominant over open competition and differentiating self-organization. For example, possibilities of global expansion and phases of shrinking and retention can be expected to alternate in complex waves (Schumpeter, [1939] 1964, pp. 175f.).

**Two Triple Helix models: an institutional and evolutionary version**

The Triple Helix of university-industry-government relations offers a rich empirical model because one can examine these relations in terms of institutional arrangements using a neo-institutional perspective, and then study collaboration patterns using, for example, social network analysis. For example, one can determine the extent to which universities are central nodes in the networks (Godin & Gingras, 2000) using centrality indicators (de Nooy *et al.*, 2005). Institutional arrangements develop historically, that is, with the arrow of time. The institutional perspective thus allows for ranges of historical case-studies on the basis of which one can develop hypotheses about the underlying dynamics of cultural evolution (e.g., Etzkowitz *et al.*, 2000).

The main advantage of using the Triple Helix model from this (neo)institutional perspective is, in my opinion, that none of the three contexts are externalized on *a priori* grounds. For example, government configurations and interventions matter for university-industry relations. Furthermore, the endogenization of techno-scientific developments into the analysis of the political economy has unavoidably led to a shift away from neo-classical assumptions about equilibrium-seeking and towards a perspective on fluxes in evolutionary economics (Nelson & Winter, 1977; Dosi, 1982). The dynamics of a knowledge-based economy are in important respects different from those of a political economy (Schumpeter, 1912; Leydesdorff, 2006b).

The alternative of an evolutionary appreciation of the Triple Helix model focuses on the three functional dynamics carried by these institutions and their networks: (*i*) wealth generation, (*ii*) novelty production, and (*iii*) political control. These three functions have, in my opinion, to be invoked and thus to operate as social coordination mechanisms in the case of innovation. Social functions, however, can be deconstructed using Luhmann's perspective (as a critique of Parsons' structural-functionalism) in terms of differently coded and symbolically generalized media of communication.

Parsons (1963a, 1963b, 1968) was the first to propose symbolic generalization, but he did not identify symbolic generalization in the codes of communication as potential drivers of the differentiation in social systems, although he understood that codes operate latently (Lazarsfeld & Henry, 1968). This latency provides the symbolically generalized codes with the status of "genotypes:" the social system is replicated in terms of coding of the expectations, wheras the "phenotypical" units of interaction can be expected to operate in terms of (institutional) agency. The "phenotypes" are necessarily historical (Hodgson & Knudsen, 2011).

As against a double helix, a triple helix is sufficiently complex to contain all these kinds of non-linear behavior. The "double helix" can first be considered as a co-evolutionary model that allows, among other things, for "mutual shaping" and thus for the emergence of historical trajectories. These trajectories may be "locked-in" into a regime when the third helix is relatively stable. For example, the constitutional stability of nation states (after the period of gestation from 1848 to 1870) allowed for the construction of national systems of innovation during the 20$^{th}$ century. Government procurement systems in the health or energy sector may weaken the market and innovation because of the imposed selection of specific technologies (Hughes, 1983). Similary, the Soviet Union, controlling the market from a centralized perspective, could not maintain its innovative edge during the Cold War.

As yet, there is no prescription for the right mix among the coordination mechanisms: China has been extremely successful after introducing its "market system" at the 14$^{th}$ Congress of the Communist Party in 1991 (Leydesdorff & Guoping, 2001). In this system, the role of government in guiding techno-economic developments is much stronger than is typical for Western democracies. Although more recently the emphasis in government funding has shifted towards "business expenditure on R&D" (BERD in OECD terminology; cf. OECD, 1976), the boom in scientific publishing during the period 1995-2005 was based on the funding of academic research (that is, "higher eduction expenditure on R&D" and "government intramural expenditure on R&D"; cf. Shelton & Leydesdorff, 2011).

Is the Triple Helix a sufficiently complex model to explain these phenomena? Why would one not add "the public" as a fourth helix, "sustainability" as a fifth, etc., as has been suggested by colleagues and in the literature (e.g., Carayannis & Campbell, 2009 and 2010). I would argue that one would expect a whole alphabet of relevant dimensions to operate in a complex system such as interhuman communications (Simon, 1973). However, three of them seem currently to be generalized symbolically to such an extent that they can globally carry the knowledge-based development of the social system.

For reasons of parsimony, in my opinion, one should be cautious in generalizing beyond the Triple Helix model to an *N*-tuple of helices so long as one is not able to operationalize and show development in the relatively simple case of three dimensions (Leydesdorff, 2012). Dimensions other than economic transactions, political power, and scientific truth-finding may yet not be sufficiently codified and globalized, but perhaps can be further developed in future stages of cultural evolution. Luhmann (2004), for example, also distinguished law as an autopoietic subsystem, along with love (Luhmann, 1986) and art (Luhmann, 2000). From a methodological perspective, the issues are already quite complex with three subdynamics, although I began to explore the working of a fourth helix (of international collaborations) in a study about Japan (Leydesdorff & Sun, 2009)—followed most recently by one about Korea (Kwon *et al*., 2012).

In the Japanese case, for example, our results suggested that despite strong government incentives and the prevailing rhetoric, over the last three decades university-industry relations have steadily declined in terms of coauthorship relations. International coauthorship, however, has steadily increased in all three helices, but particularly among academics. After the changes at the global level brought about by the demise of the Soviet Union and the opening of China to the market system in 1991, the synergy between the international relations of Japanese authors and their national Triple-Helix relations gradually increased.

In some cases, university and industrial researchers in Japan are indirectly linked by foreign coauthorship relations from which the Japanese system at the aggregated level is able to profit as a retention mechanism. We did not find such a change using Canadian data, because mobility in international relations with the USA has always been immensely important in Canada, nor for Korea, because the political direction of the Korean government controls the science system more intensively between elections (Park & Leydesdorff, 2010).

**What is the position of the university?**

Shinn (2002) noted that the strength of the Triple Helix model is its focus on university relations because the university can be expected to play a crucial role in the long-term dynamics of the knowledge-based system given the steady stream of students as influx and output to the labour market. Is the university as an institution changing because of commercial pressures, changes (e.g., cuts) in government funding, entrepreneurial activities (such as patents, licensing, spin-offs, etc.)? Is there a tendency to provide funding for elite universities as against rank-and-file state universities that focus primarily on mass education?

In a comprehensive study of university publishing comparing various rankings, Halffman & Leydesdorff (2010) found a global *decrease* of differentiation among universities. Universities increasingly seem to be involved in a global regime of competition that sets imperatives and therefore tends to lead to institutional isomorphism (Powel & DiMaggio, 1991). Leydesdorff & Meyer (2010) noted that after a long period of growth, university patenting reached a plateau and may even be on the decline since the mid-2000s.

Whether we are witnessing "the End of the Bayh-Dole Effect" as this contribution was entitled with reference to the Bayh-Dole Act of 1981 in the USA, remains to be seen given recent changes in the finance system. However, it seems to me that the main reason for these long-term changes could be the pressures of global ranking to which universities are now subjected as another selection environment. Rankings take high-quality publications into account, but usually do not attribute much weight to patenting and commercial activities.

In summary, selection environments that differ from natural selection are dynamic and can be expected to differentiate when the relevant environments become more complex. In evolutionary economics, one first distinguished between market and non-market selection environments (Nelson & Winter, 1982). This has changed evolutionary theorizing because from a biological perspective one cannot distinguish easily between natural and non-natural selection environments without leaving the Darwinian paradigm. (This discussion has led to a debate about Lamarckian evolution, but Hogdson and Knudsen (2011) have shown convincingly that this line of argument can be considered a red herring.)

The specification of different selection environments can only be done in the form of hypotheses because selection environments operate negatively and cannot be directly observed. Variation can be observed historically; the specification of a selection mechanism remains a theoretically informed hypothesis. Andersen (1994) noted that Nelson & Winter (1982) and the cohort of neo-Schumpeterian scholars following them failed to specify "what is evolving?" in their systems view. Therefore, this type of "history-friendly" modelling (Malerba *et al.*, 1999) tends to degenerate into a theory of the firm as the observable units of analysis (e.g., Casson, 1997). However, firms are carriers of the system at the phenotypical level and thus do not "evolve," but at best compete with one another.

What evolves, in my opinion, are the functionally differentiated coordination mechanisms in interhuman communication. These systems are functionally different in terms of their codes of communication. Freedom to recombine these codes is needed for innovation. A coevolutionary model of (technological) supply and (market) demand is insufficiently complex because a Triple Helix of economic exchanges, power relations, and knowledge production is involved. The third (sub)dynamics may stabilize or destabilize an equilibrium dynamics between the other two. Destabilization of non-natural selection mechanisms may lead to meta-stabilization, hyper-stabilization in a lock-in, and/or globalization. The transition from a trajectory (e.g., Mode-1) to a regime (e.g., Mode-2) is then a systems effect which can no longer be attributed to individual intentions or specific actions.

**The formal Triple Helix model**

In another branch of mathematical biology, mutual information in three dimensions was used by Ulanowicz (1986, at pp. 143f.) to develop Ascendency Theory in ecology. This information measure had been around in information theory, where it was mainly used for modelling the interactions among alphabets as different codes (Abramson, 1962) or as a non-parametric alternative to three-way interaction effects in the variance (McGill, 1954). The particularity of the measure is that it can become *negative*.

Shannon (1948) first proposed considering (and measuring) information counter-intuitively as *uncertainty*. For example, if all bets are off in a 50-50 chance process, this provides us with precisely one bit of information. Shannon-type information is by definition positive. Krippendorff (2009) showed that the mutual information in three (or more) dimensions can therefore no longer be considered as Shannon-type information. Yeung (2009) proposed to consider it as a *signed* information measure.

Since information is defined as positive uncertainty in this tradition, a negative information content would reduce uncertainty. In operation, however, overall uncertainty necessarily increases—analogously to the second law in thermodynamics (Theil, 1972)—and a reduction of uncertainty or, in other words, an increase in redundancy can only be caused by the generation of a larger maximum information content in the system (Brooks & Wiley, 1986). The maximum information content of a system is by (Shannon's) definition equal to the logarithm of the number of options available.

It can perhaps be debated whether in natural evolution the number of options continuously increases, but in cultural evolution, under the condition of functional differentiation among the codes of communication, this number proliferates because the same events can be provided with different meanings. Furthermore, as noted, meanings refer to horizons of meaning which can also be expanded over time. The expansion of horizons of meaning is generated by next-order models of the meanings constructed hitherto. The techno-sciences (e.g., ICT) are powerful drivers in generating new meaning that, if successful, can be used for innovating systems that were previously considered as "naturally given". Technological developments may outmode and overwrite previous generations of technology. Thus, an evolutionary pump of redundancy is generated by the self-organization of the techno-sciences.

Similarly, the coding of transactions in the economy in terms of money, credit, stocks, etc. adds layers to the possibilities of economic exchanges. The *trias politica* enriches political and legal discourse with dimensions of potentially different meanings. In general, the proliferation of fantasies and wishes grows faster than the options that can historically be realized (Weinstein & Platt, 1969). The crucial question becomes: what can be retained as synergy—at the social level!—from these interactions among expectations? How should the institutional make-up of a knowledge-based system be arranged reflexively?

In the end, these are also normative questions, but the analytical angle of the mutual information in three dimensions provides us with an access that can inform the discourse. It could be shown (Krippendorff, 2009; Leydesdorff, 2010a) that this potentially negative measure provides us with a *difference*; namely, the difference between the historically generated (Shannon-type) interaction information and the redundancy—that is, the number of options—generated in loops of reflexive (interhuman) communication. This difference can either add to the uncertainty in the historical system or reduce the prevailing uncertainty, in accordance with the sign of the measure.

Let me offer the simple example of a family consisting of a mother, a father, and a child. If there is synergy in this family system, the child may address a question either to the father or to the mother, but the answers will not be confusing. One can in such a case assume an *a priori* synchronization between the possible answers of both parents. The child has no access to this synchronization, but this latent structure reduces the prevailing uncertainty. Similarly, in the case of a divorce this system may break down and the structure in the communication may become complex and confusing.

Analogously, in university-industry-government relations: structural relationships between any two partners can emerge and can be dissolved. Is there synergy among the fluxes of communication carrying differently codified meanings, and can this be measured as a reduction of prevailing uncertainty at the systems level? Let me hasten to say that this is a longer-term research program that is not yet fully understood in terms of all the methodological problems involved (e.g., Leydesdorff, 2010a; De Nooy & Leydesdorff, 2009), including its status in relation to possible alternatives such as measuring interaction effects in three-way logit models. Using different proxies, however, we have applied this model to the Dutch, German, and Hungarian innovation systems and the results have been fruitful in terms of heuristics.

For example, in the Dutch system this synergy can be found at the national level, whereas in Germany it can be found at the level of the federal states (Leydesdorff & Fritsch, 2006; Leydesdorff *et al.*, 2006). In Hungary, the system has lost its systemic character at the national level: after the changes of the 1990s the western part was increasingly integrated into the dynamics of Western Europe; greater Budapest has become a metropolitan innovation system; and the eastern part is still controlled by the dynamics of the old state-led innovation system (Lengyel & Leydesdorff, 2011). In a recent study (Strand & Leydesdorff, in preparation) yet another pattern was visible in Norway.

The Triple Helix indicator is available as a computer program at http://www.leydesdorff.net/th2/th.exe. One can play with the program and see in which configurations the *T*-values (*T* for transmission) in three dimensions become negative and/or positive. This adds to my argument that the synergy from university-industry-government relations is based on specific trade-offs between expectations and institutionalizations, and should not be answered by sweeping statements about the "enterpreneurial university" or the increased threat of commercialization. Such idealizations exaggerate by using ideal-typical metaphors.

Indeed, the use of such claims, for example, in public relations, may indicate a problem. For example, in the Netherlands, the Technical University of Twente (in Enschede) advertises itself as "entrepreneurial" whereas various indicators show that the Technical University in Eindhoven may have a stronger record in this respect. The structural aspects of the Einhoven region favour such a role, whereas this synergy has been absent in Twente (Leydesdorff *et al.*, 2006). Using patent analysis—perhaps, to be considered as a proxy for commercialization—Leydesdorff & Bornmann (2012), for example, found Enschede only in the top-25% of Dutch cities, wh ereas Eindhoven had by far the largest score among the cities in the top-1%.

**Interpenetration, interaction, and interfacing**

Commercialization can be (*i*) ad hoc in interactions between academia and industry, (*ii*) structural in the organization at interfaces of a knowledge-based arrangement, and (*iii*) it may affect the cognitive core of the scientific development or remain mainly a condition. The concern is mainly about the third category: Do interactions and interfaces lead to interpenetration of commercial considerations into the process of science itself and if so to what extent? Can they become so dominant that scientific disciplines or specialties are locked-in?

Unlike Parsons, who used "interpenetration" as a term for any cross-cutting between cultural expectations and social structures (e.g., Parsons & Platt, 1973, at p. 36), Luhmann reserved this word for the reflexive coupling that is additionally possible because of reflexivity between the structurally coupled layers of interhuman communication and human consciousness. In his model, the performatively functional subsystems of communication (the economy, science, politics, etc.) do not interpenetrate one another, but merely disturb one another in what Luhmann, Ziegert, & Kastner (2004, in my opinion, erroneously)[3] considered as yet another structural coupling.

The social system operates in terms of translations among its subsystems which can each be expected to perform their own autopoiesis (under the condition of modernity). Luhmann (1990, at pp. 635 ff.) provided an amazingly precise specification of this mechanism, but he labeled it differently (cf. Leydesdorff, 2010). Let me cut this discussion short by summarizing my own model: the symbolically generalized codes of communication span the social system in perpendicular directions, like in an orthogonally rotated factor model. However, this remains an analytical reconstruction of the social constructs that can be entertained implicitly or explicitly by both analysts and the participating agents who exist under historical conditions.

The observables are contained in a historical "life-world". As with factor loadings or factor scores, manifestations are never "pure" factors, but always based on trade-offs. The analyst (as against the participant) may be able to claim additional knowledge on the basis of his or her particular perspective, but scholarly discourse develops as an evolutionarily next-order mechanism of communication (for example, in the "context of justification") that emerges from and loops into the local level. All codes of communication can historically be interfaced and interact at the level of organizations.

---

[3] In my opinion, the functional subsystems are not structurally coupled, but structurally part of the social system and therefore operationally coupled, namely, coupled whenever the social system operates.

When further formalized—that is, as decisions are increasingly taken according to interpersonal criteria and decision rules—organization itself can also be developed as yet another symbolically encoded subsystem of communication (Achterbergh & Vriens, 2009; Leydesdorff, 2006, at pp. 149 ff.). Its historicity, however, remains as a footprint in the further development of interhuman communication systems. These institutional footprints enable us to retain wealth from knowledge.

Finally, "interaction" can be a confusing concept when the interacting units are not sufficiently specified. In *this* context of organization and self-organization of the expectations in communication systems, interactions among expectations can be considered as generating the variation in this system. Variation is crucial, for example, for making the system more complex and hence eventually more knowledge-based. From this perspective, the empirical question is whether and to what extent commercialization can be expected to provide us with new opportunities to raise new scientific questions, articulate demand, etc.?

**Conclusions**

I hesitated to call my perspective a "sociological" one. Perhaps "communication-theoretical" might be better because of the use of this perspective in Luhmann's sociology of communication and Shannon's mathematical theory of communication. Unfortunately, large parts of sociology, particularly in Europe (and notably in STS), have abandoned the structural perspective because structure has been associated with reification (Ashmore, 1989; Latour, 1988; Woolgar, 1988). I have argued that the latent functions that can be hypothesized can also be deconstructed as interhuman communication systems operating with a wealth of potential codifications.

The interhuman communication of meaning (as different from and additional to the communication of information) provides events with predicates that are culturally constructed (Husserl, 1929). Sociologists have begun to explore this new domain empirically (Berger & Luckmann, 1966; Schütz, 1932), theoretically (Luhmann, 1995a), and methodologically (Leydesdorff, 2010a and c, 2011). The interactions among the different codes operating in economic transactions, political power relations, and scientific validation have been leading me further to develop a possible explanation for a knowledge-based economy (Leydesdorff & Zawdie, 2010).

How does this research programme relate to the concern expressed in the title of this topical issue, namely, "the philosophy of commercialized science?" In my opinion, this claim can itself perhaps be considered as a social text. I envisage that others in this collection may address the debate about Newton's position ignited by Boris Hessen's (1931) famous challenge to show the material interests in Newton's working life (Basalla, 1968). In action and historical organization the different codes of communication are always historically integrated in a specific organizational format. This integration may be suffocating—as it probably was not in the case of Newton, but most likely was in the case of Lysenko (Graham, 1974; Lecourt, 1976).

Codifications compete evolutionarily as "genotypes" in interhuman communications, but all "genotypes" can be expected to remain available during the replication. Human agents or institutional agents (e.g., firms) are not the evolving units in cultural evolution, but our communication structures are. One would not be able to pay with a credit card in ancient Rome, but perhaps one could understand the Latin. The Latin has remained the same, but the codification of economic transactions was further developed.

The commercialization of the sciences can be considered as the phenotypical result of interactions, interfacing, and interpenetration among these differently coded communications in historical configurations (Hessels *et al.*, 2011). One can expect the synergetic effects of these interactions, interfaces, and interpenetrations to differ among sciences, regions, countries, etc. Thus, the themes of university-industry-government relations and the (potential) effects of commercialization offer a rich domain for empirical research. For example, at some places (e.g., in Eastern Hungary) more opening to the markets may be desirable, while at other places the configurations of interests tend to drive the organizational dynamics into a lock-in as a hyper-stabilization instead of allowing for meta-stabilization and self-organization of communications as a concurrently possible dynamic (Dolfsma & Leydesdorff, 2009; Leydesdorff & Zawdie, 2010).

**Acknowledgements**
I am grateful to Iina Hellsten, Laurens Hessels, Caroline Wagner, and Sally Wyatt for comments on a draft version of this paper.